\documentclass[twocolumn,superscriptaddress,showpacs,nofootinbib,notitlepage]{aastex631}
\usepackage{graphicx,subfigure,amsmath,amsfonts,amssymb,multirow,mathrsfs,ulem,threeparttable}
\usepackage{hyperref}
\usepackage{epstopdf}
\begin{document}
\title{Modeling the JWST high-redshift galaxies with a general formation scenario and the consistency with the $\Lambda$CDM model}

\correspondingauthor{Yi-Zhong Fan}
\email{yzfan@pmo.ac.cn}

\author[0000-0003-1215-6443]{Yi-Ying Wang}
\affiliation{Key Laboratory of Dark Matter and Space Astronomy, Purple Mountain Observatory, Chinese Academy of Sciences, Nanjing 210033, People's Republic of China}
\affiliation{School of Astronomy and Space Science, University of Science and Technology of China, Hefei, Anhui 230026, People's Republic of China}

\author[0000-0003-4631-1915]{Lei Lei}
\affiliation{Key Laboratory of Dark Matter and Space Astronomy, Purple Mountain Observatory, Chinese Academy of Sciences, Nanjing 210033, People's Republic of China}
\affiliation{School of Astronomy and Space Science, University of Science and Technology of China, Hefei, Anhui 230026, People's Republic of China}

\author[0000-0002-4538-8526]{Guan-Wen Yuan}
\affiliation{Theoretical Physics Division, Institute of High Energy Physics, Chinese Academy of Sciences, Beijing 100049, People’s Republic of China}
\affiliation{Key Laboratory of Dark Matter and Space Astronomy, Purple Mountain Observatory, Chinese Academy of Sciences, Nanjing 210033, People's Republic of China}

\author[0000-0002-8966-6911]{Yi-Zhong Fan}
\affiliation{Key Laboratory of Dark Matter and Space Astronomy, Purple Mountain Observatory, Chinese Academy of Sciences, Nanjing 210033, People's Republic of China}
\affiliation{School of Astronomy and Space Science, University of Science and Technology of China, Hefei, Anhui 230026, People's Republic of China}

\newcommand{\ud}{\mathrm{d}}
\begin{abstract}
Early results from the James Webb Space Telescope (JWST) observations have hinted at two traces beyond the standard cosmological framework. One is the extraordinarily high stellar masses and their density at $z=7.5\sim9.1$, another is the unexpected abundance of ultraviolet (UV) bright galaxies at $z\ge10$. Nevertheless, both pieces of evidence are not statistically robust yet. In this work, we construct rest-frame UV luminosity functions (LFs) based on a general formation model for these high-redshift galaxy candidates, since UV LFs always carry the information of stellar formation efficiency (SFE), initial mass function (IMF), dust attenuation, and other crucial elements for galaxy evolution. By updating the massive galaxies candidates with spectroscopic observations and exploring the parameter space of SFE, we are able to reasonably explain the cumulative stellar mass density within the redshift range of $7.5\sim9.1$, with only one galaxy exhibiting unusual characteristics. We also reveal a potential nonmonotonic trend of SFE with the increasing redshift.  At higher redshift ($z\sim13$), bright UV LFs can be well fitted with non-dust attenuation or top-heavy IMF for Population III stars. The Population III star scenario can also naturally account for the possible dip of the peak SFE evolution curve at $z\sim9$.
\end{abstract}

\section{Introduction}\label{sec:1}
The successful performance of the James Webb Space Telescope (JWST) enables human beings to glimpse at the early universe and even unveil the first generation of the galaxies and stars. To date, photometric measurements have identified some candidates for the very-high-redshift galaxies up to  $z\sim16$ \citep{2023arXiv230204270A, 2023MNRAS.523.1009B, 2023MNRAS.518.6011D, 2023ApJ...946L..13F, 2023ApJS..265....5H, 2023arXiv230202429P}, with a spectroscopic redshift confirmation at $z\sim13$ \citep{2023NatAs...7..622C, 2023NatAs...7..611R}. This breaks the record of $z\sim11$ \citep{2016ApJ...819..129O,2021NatAs...5..256J}  based on observations from the Hubble Space Telescope (HST).

Notably, current observations present intriguing features that may challenge the standard cosmological model. One of them is the discovery of several massive galaxies with $\log M_{*} > 10 \, M_{\odot}$ from $z=7.4$ to $z=9.1$  \citep{2023Natur.616..266L}. Another is the greater numbers of bright $z \gtrsim 10$ galaxies candidates than most theoretical models predicted, as indicated by the rest-frame UV luminosity function \citep[LF;][]{ 2023arXiv230413721A, 2023MNRAS.523.1009B, 2023MNRAS.523.1036B, 2023MNRAS.518.6011D, 2023arXiv230406658H, 2023arXiv230414469M, 2023ApJ...946L..35M, 2022ApJ...940L..14N, 2023arXiv230202429P}.

\citet{2023NatAs.tmp...77B} claimed that there is a robust discrepancy between the observed high-redshift galaxy candidates and the $\Lambda$CDM model. To account for the very high stellar mass density observed by JWST \citep{2023Natur.616..266L}, the star formation efficiency (SFE) is required to be larger than 0.57 at $z\approx7.5$ and even close to 1 at $z\approx9.1$ in the standard cosmological framework, which are beyond the value from both previous empirical models \citep{2015ApJ...813...21M, 2018MNRAS.477.1822M, 2019MNRAS.488.3143B} and observational findings  \citep{2022ApJS..259...20H}. Only some special cooling mechanisms, such as the feedback-free starbursts \citep{2023MNRAS.523.3201D}, can significantly enhance the SFE by suppressing feedback from stars and supernovae, thus preventing disruptions in the star-forming process. \citet{2023MNRAS.518.2511L}, \citet{2023arXiv230607781F} and \citet{2023arXiv230100347W} also highlight a tentative tension between the formation of these high-redshift massive galaxy candidates and $\Lambda$CDM model. For higher-redshift ($z \gtrsim 10$) candidates, UV LFs shed valuable 
lights on the first generation of stellar populations, star formation activity, galaxy evolution, and other crucial elements. However, whether the conventional dark-matter (DM) -driven galaxy formation paradigm is already falsified has turned into a controversy \citep{2022ApJ...939L..31H}. Based on the classic Sheth-Tormen mass function \citep{2001MNRAS.323....1S} for DM halos, \citet{2022ApJ...938L..10I} found that the upper bounds of galaxy LFs (without dust attenuation) are lower than the observations from JWST, unless a high SFE or a top-heavy initial mass function (IMF) for the stellar population is assumed. \citet{2023arXiv230404348Y} drew a similar conclusion using a well-established semianalytic model. It should be noticed that such high SFE can explain the bright UV LFs and high cumulative stellar mass simultaneously, but still need more verification to confirm its rationality. On the contrary, using cosmological simulations, \citet{2023arXiv230715305S} concluded that burst star formation history can explain the bright-end UV LFs naturally without invoking other modifications. \citet{2023MNRAS.522.3986F} and \citet{2023MNRAS.521..497M} argued that a minimal amount of dust could significantly enhance the brightness of high-redshift galaxies, thereby reconciling the LFs with observations. 

In this work, we adopt a general formation model to explain the massive galaxy candidates and the bright UV LFs. First and foremost, we recalculate the number density from \citet{2023Natur.616..266L}, considering the latest spectroscopic confirmations. Up to now, three out of the total reported massive galaxies from \citet{2023Natur.616..266L} have been identified with lower spectroscopic redshifts. IDs of 13050 ($z\sim8.14$, $\log M_{*} \sim 10.14 \, M_{\odot}$), 35300 ($z\sim 9.08$, $\log M_{*}\sim 10.40 \, M_{\odot}$), 39575 ($z\sim8.62$, $\log M_{*} \sim 9.33 \, M_{\odot}$) mentioned in \citet{2023Natur.616..266L} have been confirmed as CEERS 3210 \citep[$z\sim5.624$,][]{2023arXiv230200012K}, CEERS3\_1748 \citep[$z\sim7.769$, $\log M_{*}\sim 9.49 \, M_{\odot}$,][]{2023ApJ...949L..25F} and CEERS1\_3910 \citep[$z\sim7.99$, $\log M_{*}\sim 8.94 \, M_{\odot}$][]{2023ApJ...949L..25F}, respectively. 
The corresponding cumulative stellar mass density decreases due to the exclusion of one massive candidate $\log M_{*} > 10 M_{\odot}$ in each redshift bin ($7<z<8.5$ and $8.5<z<10$). Consequently, the tension between JWST and HST+IRAC \citep{2021ApJ...922...29S} can be mildly alleviated. Considering both of these anomalous observations can be explained by a high SFE, we explore the parameter space and find a gradual evolution of SFE from $z\sim 4$ to $\sim13$, which is consistent with previous works \citep{2018MNRAS.477.5382B, 2018MNRAS.477.1822M} suggesting that the profile of SFE-$M_{\rm h}$ will change at different redshifts. To analyze the rest-frame UV LFs and the cumulative stellar mass density, we take into account the influences of dust attenuation and cosmic variance (CV). 
Throughout this work, we assume the Planck flat $\Lambda$CDM cosmology with $H_0=67.36 \rm \, km \, Mpc^{-1} \, s^{-1} $, $\Omega_{\rm m}=0.3153$, and $\Omega_{\Lambda}=0.6847$ \citep{2020A&A...641A...6P}. All magnitudes are in the absolute bolometric (AB) system \citep{1983ApJ...266..713O}. 

\section{Method}\label{sec:2}
We construct the UV LFs from $z\sim4$ to $z\sim13$ based on a general model of galaxy evolution. For $z\le10$, the DM halos evolve following the Sheth-Mo-Tormen function \citep{2001MNRAS.323....1S} that a spherical region collapses when the initial density exceeds a certain threshold value. For $z\ge10$, \citet{2007MNRAS.374....2R} obtained the mass function $f(\sigma)$ of DM halos using the high-resolution N-body simulations. The number of DM halos per unit mass per unit comoving volume can be written as 
\begin{equation}\label{eq:1}
	\frac{{\rm d} n}{d \ln M_{\rm h}} = M_{\rm h} \cdot \frac{\rho_0}{M_{\rm h}^2} f(\sigma) \, \bigg \vert \frac{{\rm d} \ln \sigma}{{\rm d} \ln M_{\rm h}} \bigg \vert ,
\end{equation}
where $\rho_0$ is the mean density of the Universe and $\sigma$ is the rms variance of mass, which is determined by the linear power spectrum and the top-hat window function. The linear power spectrum can be computed using the transfer function provided by the public Code for Anisotropies in the Microwave Background  \citet[CAMB;][]{2000ApJ...538..473L}. The relevant calculation about the DM halos can be carried out with the public code HMFcalc \citep{2013A&C.....3...23M}.

Once the DM halo mass $M_{\rm h}$ and a specified redshift are defined, the mass growth rate $\dot{M_{\rm h}}$ can be obtained from N-body simulation. Interestingly, the calculation of $\dot{M_{\rm h}}$ is conveniently achievable using Eq.~(B1)-(B8) provided by \citet{2015ApJ...799...32B}, and these equations offer highly accurate representations by matching simulation results with analytic formulas. In \citet{2015ApJ...799...32B}, high-redshift galaxy evolution is constructed within a wide range of redshift ($4<z<15$) since the halo properties are obtained primarily from Bolshoi simulation results \citep{2011ApJ...740..102K}, encompassing the redshift from $z=80$ to the present. This evolution spans a diverse range of halo masses, extending from approximately $10^9 M_{\odot}$ to about $10^{15} M_{\odot}$ at different redshifts. As shown in their Fig.~18, the parameterized halo mass accretion rates are compared with the Bolshoi simulation \citep{2011ApJ...740..102K}, the 
Bolshoi-Planck simulation (i.e., the
Bolshoi simulation with the Planck best-fit cosmological parameters \citep{2014A&A...571A..16P}), and the Lb125 simulation 
\citep{2005MNRAS.364.1105S}
across several characteristic halo masses, revealing precise deviations between the analytic formulas and the simulations. The accretion rate of baryon is $\dot{M_{\rm b}} = f_{\rm b} \times \dot{M_{\rm h}}$, where $f_{\rm b}$ is the cosmic baryon fraction defined as $f_{\rm b}\equiv \Omega_{\rm b}/\Omega_{\rm m} =0.156$. Further, the star formation rate (SFR) is ${\rm SFR}= f_{*} \times \dot{M_{\rm b}}$, where $f_{*}$ is SFE that we concentrated on.  We employ an adjustable equation for $f_{*}$ and consider its variation at different redshifts. $f_{*}$ is given by \citep {2010ApJ...710..903M}:
\begin{equation}\label{eq:2}
	f_{*}=\frac{2\epsilon_{\rm N} }{\bigl( \frac{M_{\rm h}}{M_1} \bigr )^{-\beta} + \bigl( \frac{M_{\rm h}}{M_1} \bigr)^{ \gamma}},
\end{equation}
where $\epsilon_{\rm N}$ is the normalized constant, $M_1$ is the characteristic mass where the value of SFE is equal to $\epsilon_N$, and $\beta$ and $\gamma$ are slopes that determine the decrease in SFE at low and high masses, respectively.

With \autoref{eq:2}, SFE exhibits an initial increase followed by a subsequent decrease with the increasing mass of DM halos. Such profile of $f_{*} - M_{\rm h}$ evolution was obtained by \citet{2007ApJ...667..760Z} for the first time. Within the framework of the halo occupation distribution, they calculated the SFE at $z\sim0$ (Sloan Digital Sky Survey) and $z\sim1$ (DEEP2 Galaxy Redshift Survey). Their analysis revealed that the SFE reaches the maximum at $M_{\rm h} \sim 5\times 10 ^{11} h^{-1} M_{\odot}$ and $10\times 10^{13} h^{-1} M_{\odot}$ for the two respective redshifts. Analogously, \citet{2006MNRAS.368....2D} compared the compression rate and cooling-loss rate to derive the shock stability criterion. They concluded that the star formation would be efficient with $M_{\rm h} \le 10^{12} M_{\odot}$ because enough cold gas filaments feed the DM halo, which was not heated by the virial shock at that time. Moreover, their Fig. 8 illustrated the strength of different feedback processes (e.g. supernova feedback, UV-on-dust feedback, active galactic nucleus [AGN] feedback, and others) with DM halo mass increasing, indicating a minimum feedback efficiency at $M_{\rm h} \sim 10^{12} M_{\odot}$, which might be the potential candidate value of $M_{1}$. In many works, the values of $\epsilon_{\rm N}$, $M_1$, $\beta$, and $\gamma$ tend to differ. For instance, \citet{2023arXiv230505679S} gave $\epsilon_{\rm N}=0.1$, $M_1=10^{12} M_{\odot}$, $\beta=0.6$, and $\gamma=0.5$ at $z\ge9$. \citet{2022ApJS..259...20H} gave $\epsilon_{\rm N}=0.21$, $M_1=10^{11.5} M_{\odot}$, $\beta=1.2$, and $\gamma=0.5$ at $z=7$. \citet{2018MNRAS.477.1822M} gave $\epsilon_{\rm N}=0.24$, $M_1=10^{12.1} M_{\odot}$, $\beta=1.3$, and $\gamma=0.6$ at $z=8$. Because previous works suggested a tight range for $M_{1}$ and both \citet{2023arXiv230505679S} and \citet{2022ApJS..259...20H} kept $M_1$ being constant rather than a free parameter when analyzing UV LFs, we fix $M_1 = 10^{12} M_{\odot}$ in the discussion below. Besides, we test the hypothesis that treating $M_1$ as a free parameter (e.g. a Gaussian prior) and find an indistinctive impact on the posterior values of our primary concern, $\epsilon_{\rm N}$. A brief description can be found in \autoref{sec:6}. Despite the variations in these parameter values, the overall trend of the SFE evolution remains consistent. For lower-mass halos, there is not much cold gas for star formation because of the supernova feedback. And for higher-mass halos, SFE is also suppressed by the high virial temperature because of the strong negative AGN feedback. 

Following \citep{2014ARA&A..52..415M}, the UV LFs can be derived from
\begin{equation}
	{\rm SFR_{UV}}(M_{\odot} {\, \rm yr^{-1}})={\mathcal K}_{\rm UV} \times L_{\rm UV}(\rm erg \, s^{-1}\, Hz^{-1}),
\end{equation}
where ${\mathcal K}_{\rm UV} =1.15\times10^{-28}$ is the conversion factor with the Salpeter \citep{1955ApJ...121..161S} IMF at 1500 {\r A}. For lower metallicity or larger IMF, the value of ${\mathcal K}_{\rm UV}$,  which contributes to the brightening of galaxies for a fixed $ \rm SFR$, decreases (Tab.~1 in \citet{2022ApJ...938L..10I}). In that case, it suggests the observed possibilities of the existence for Population III. In addition, dust attenuation diminishes the absolute magnitude ($M_{\rm UV}$) and visibly suppresses the UV LFs. Based on the attenuation-UV slope ($\beta_{\rm UV}$) relation \citep{1999ApJ...521...64M}, we assume $\beta_{\rm UV}-M_{\rm UV}$ relations following \citet{2014ApJ...793..115B} for $z<8$ and adapt the latest results \citep{2023MNRAS.520...14C} obtained from JWST for higher-redshift galaxies. Therefore, the intrinsic $M_{\rm UV}$ can be converted into the dust-corrected $M_{\rm UV}$, being closer to the real detection. Hereto, there are equations about converting the DM halos to $M_{\rm UV}$ at arbitrary redshift. The UV LF can be written as
\begin{equation}\label{eq:4}
	\Phi(M_{\rm UV})=\phi(M_{\rm h}) \, \bigg \vert \frac{{\rm d} M_{\rm h}}{{\rm d} M_{\rm UV}} \bigg \vert,
\end{equation}
where $\phi_{M_{\rm UV}}$ is from \autoref{eq:1}, $\big \vert \frac{{\rm d} M_{\rm h}}{{\rm d} M_{\rm UV}} \big \vert$ is the Jacobi matrix mapping from $\phi(M_{\rm h})$ to $\Phi(M_{\rm UV})$, and $M_{\rm UV}$ is the dust-corrected or intrinsic magnitude, depending on whether the dust-attenuation effect is considered or not. Fitting \autoref{eq:4} to the observed UV LFs spanning a wide range of redshift, we can obtain the best-fit value for \autoref{eq:2}. Since some observed data have asymmetrical uncertainties, we utilize asymmetric Gaussian distribution as the likelihood at each redshift,
\begin{equation}
	{\rm Likelihood}=\prod^{N}_{i} {\rm AN}(f(x_i)-y_i | c_i, d_i),
\end{equation}
where AN is the asymmetric normal distribution \citep[i.e., the Eq.~(14) of][]{2013ApJ...778...66K}, ($x_i$, $y_i$) are the observed UV LFs, $f(x_i)$ is the value predicted by the UV LF model at $x_i$, $c_i$, and $d_i$ represent the deviation and the skewness, which can be obtained from the up-down errors \citep{2013ApJ...778...66K}. The prior distributions of the parameters in \autoref{eq:2} are summarized in \autoref{Tab:1}. Balancing the accuracy and the efficiency, we use the nested sampling method and adopt {\tt Pymultinest} as the sampler in the Bayesian parameter analysis. During the analysis process, we utilize 2000 live points and configure evidence tolerance $=0.5$ as the stopping criteria. 
Moreover, further tests employing 500, 1000, and 4000 live points always reveal identical posterior distributions and consistent Bayesian evidence at each redshift, suggesting the convergence of our results strongly.

\begin{table}
	\begin{ruledtabular}
		\caption{Prior distributions of the parameters for SFE}\label{Tab:1}
		\begin{tabular}{lcc}
			Parameters  &Priors of Parameter Inference    \\ \hline            
			$\epsilon_{\rm N}$    &Uniform(0.001, 1)\\
			$\beta$                  &Uniform(0.1, 2)\\
			$\gamma$             &Uniform(0.1, 2)
		\end{tabular}
	\end{ruledtabular}
\end{table}

The comoving stellar mass density contained within galaxies more massive than $M_{*}$ is \citep{2023NatAs.tmp...77B}
\begin{equation}\label{eq:6}
	\rho_{*}(>M_{*},z)=\int_{M_{\rm h}}^{\infty} {\rm d}M \, M \cdot f_{b} \cdot f_{*} \cdot \frac{{\rm d}n(M,z)}{{\rm d}M},
\end{equation}
where $ \frac{{\rm d}n(M_{\rm h},z)}{{\rm d}M_{\rm h}}$ is from \autoref{eq:1}, and $f_{*}$ is the fitting result. However, the galaxy number densities observed in finite volume always have uncertainty due to CV. Especially in small survey areas and for massive high-redshift galaxies, the effect of CV is nonnegligible and even can be comparable to the actual observations.  Therefore, we incorporate its contribution into \autoref{eq:6} and compare the cumulative stellar mass density observed from \citet{2023Natur.616..266L} with some updates. The calculation of CV follows \citet{2011ApJ...731..113M}.

\section{Result}\label{sec:3} 
\begin{figure*}[ht!]
	\centering
	\subfigure{
		\includegraphics[width=0.45\textwidth]{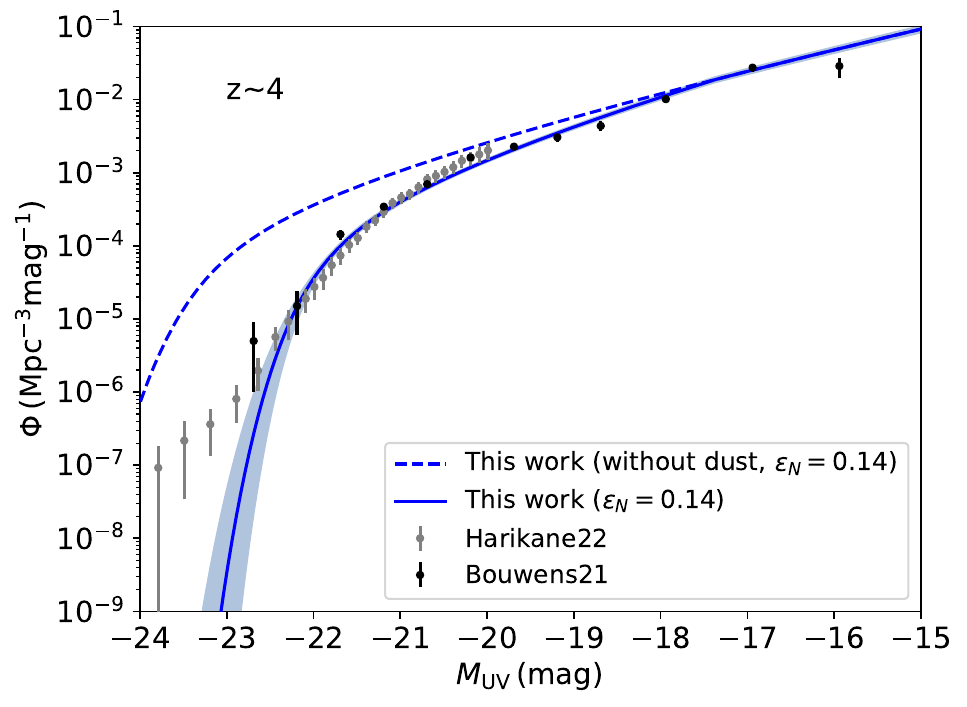}
		\hspace{12mm}}
	\subfigure{
		\includegraphics[width=0.45\textwidth]{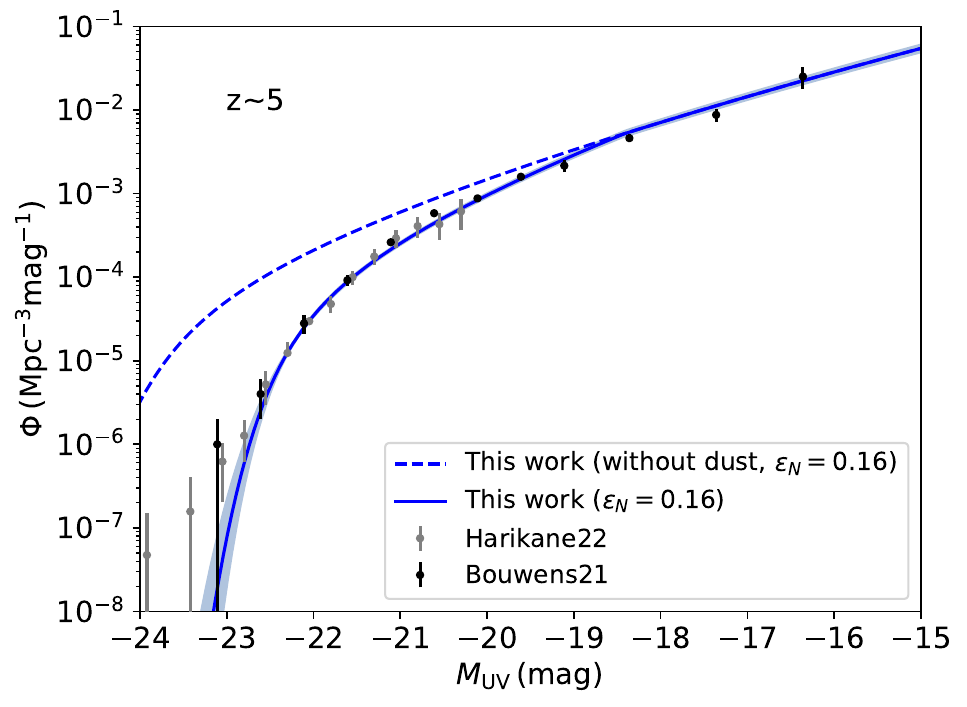}}
	\subfigure{
		\includegraphics[width=0.45\textwidth]{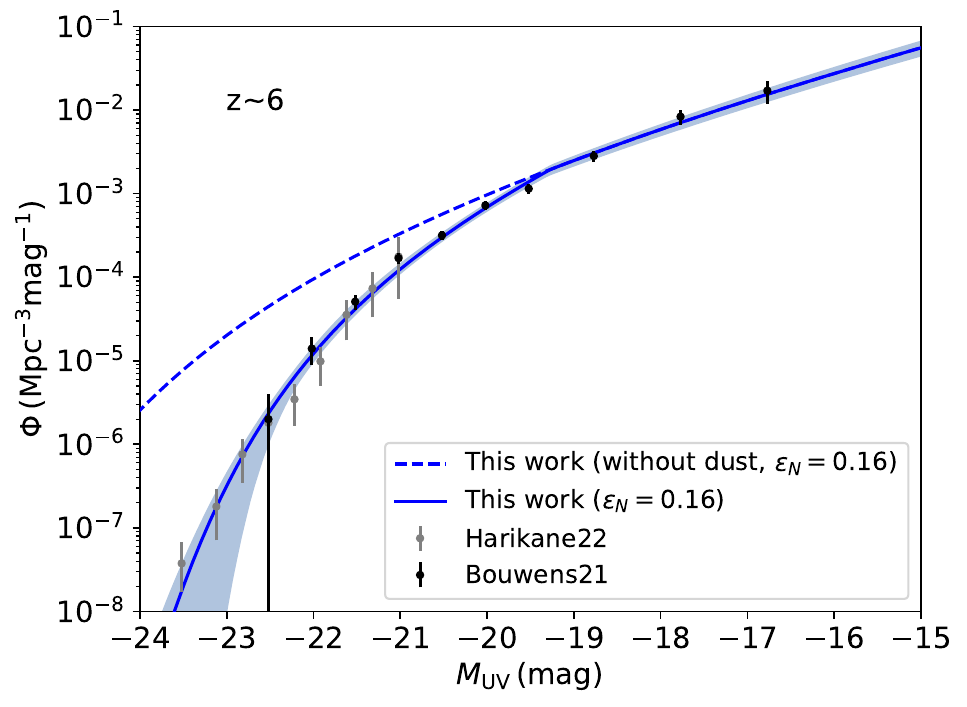}
		\hspace{12mm}}
	\subfigure{
		\includegraphics[width=0.45\textwidth]{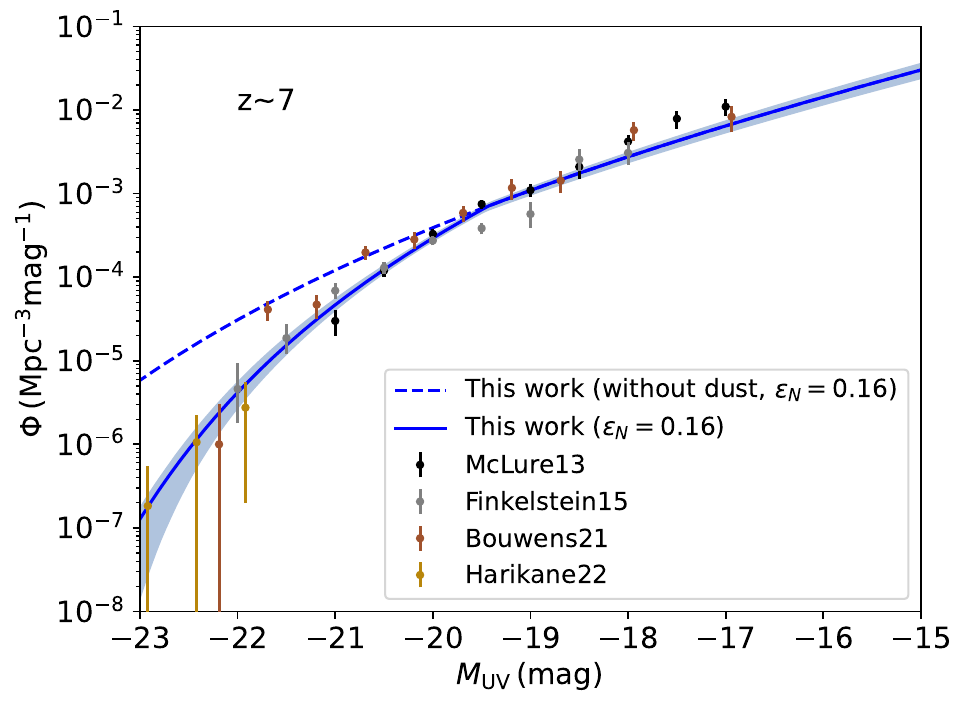}}
	\subfigure{
		\includegraphics[width=0.45\textwidth]{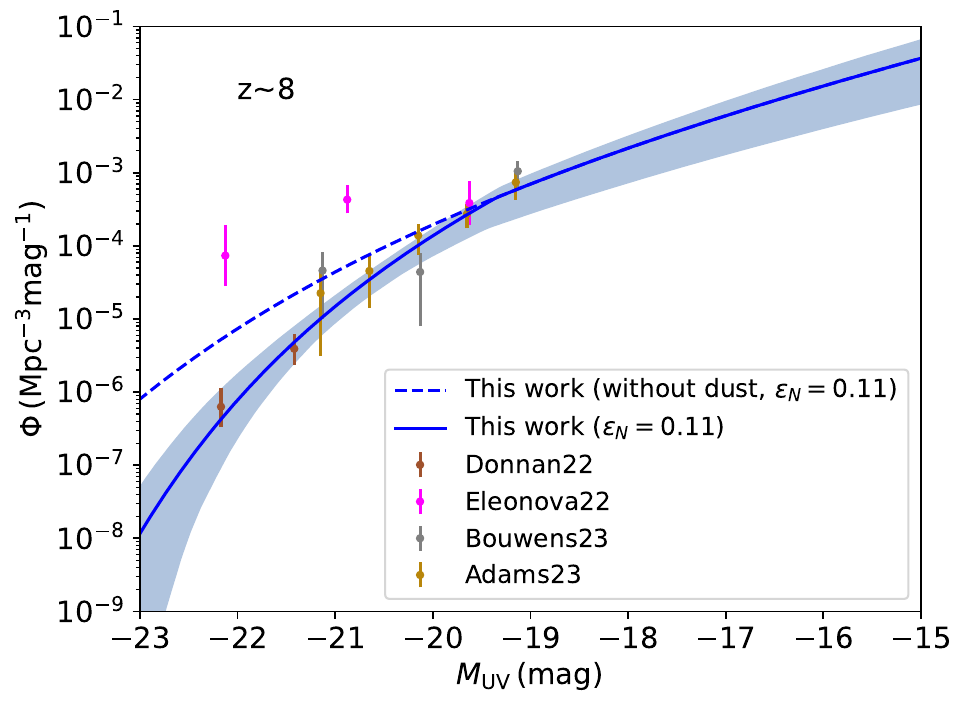}
		\hspace{12mm}}
	\subfigure{
		\includegraphics[width=0.45\textwidth]{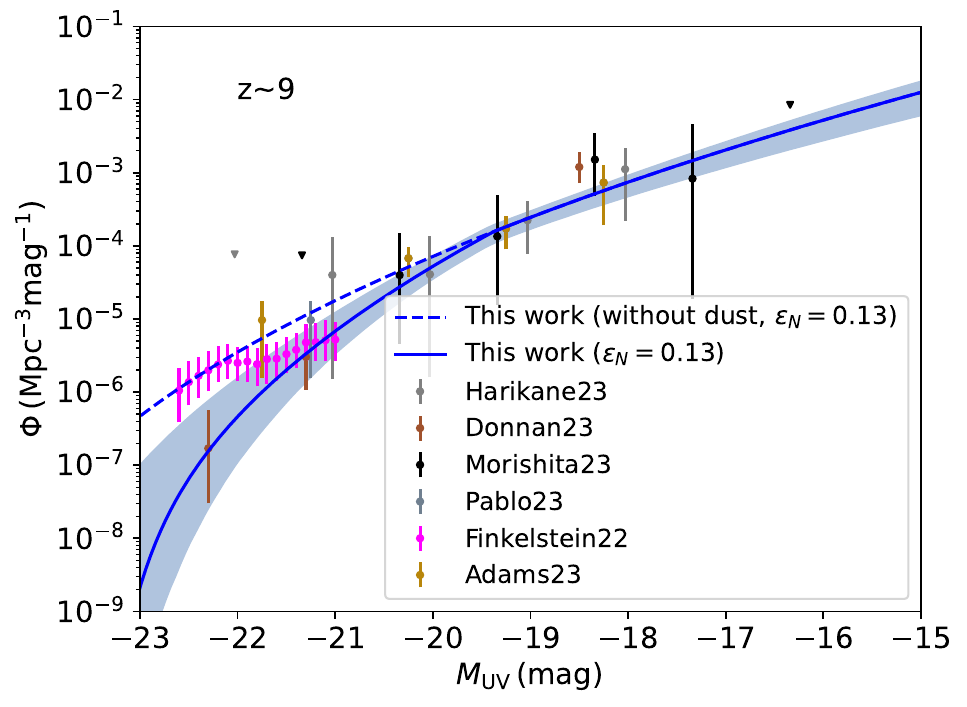}}
	
	\caption{\small The fitting results of the observed UV LFs at the redshift range of $4\le z \le 9$. The selected JWST galaxies are taken from  \cite{2013MNRAS.432.2696M, 2015ApJ...810...71F, 2021AJ....162...47B, 2022ApJS..259...20H, 2023arXiv230413721A, 2023MNRAS.523.1009B, 2023MNRAS.518.6011D, 2023ApJS..265....5H, 2023ApJ...946L..35M, 2023arXiv230202429P}. The solid blue lines depict the best fit to the entire dataset. The light blue regions represent the corresponding $90\%$ credible regions. The dashed lines illustrate the best fit of UV LFs but reshaped to the case without dust attenuation. The observational data points are marked in circles with error bars, and the upper limits are shown in lower triangles.}
	\label{fig:1}
\end{figure*}

\begin{figure*}[ht!]
	\centering
	\subfigure{
		\includegraphics[width=0.45\textwidth]{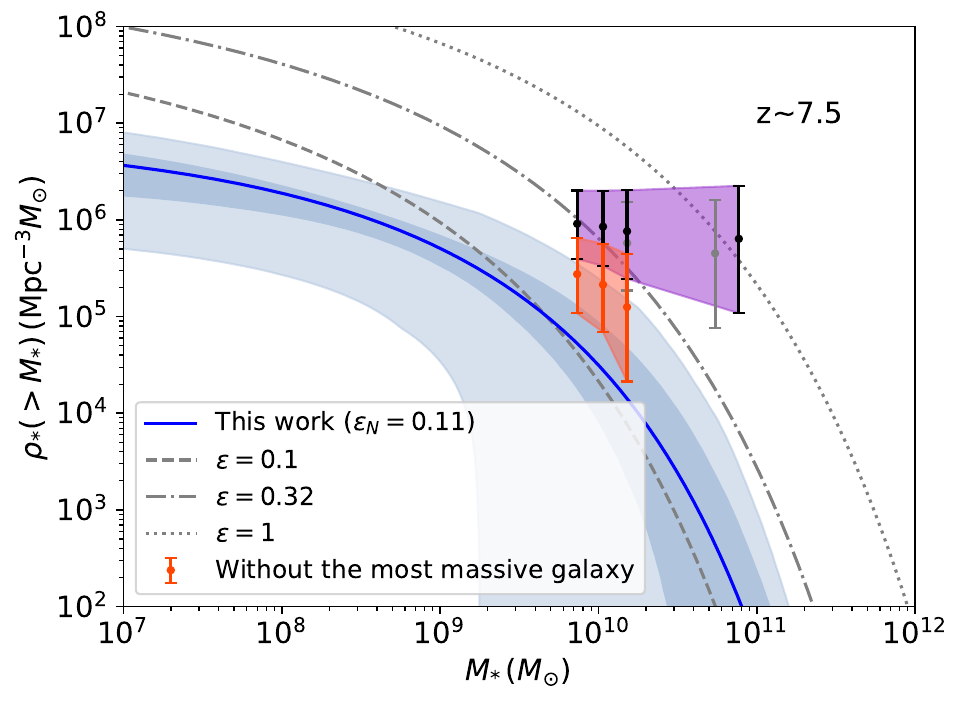}
		\hspace{12mm}}
	\subfigure{
		\includegraphics[width=0.45\textwidth]{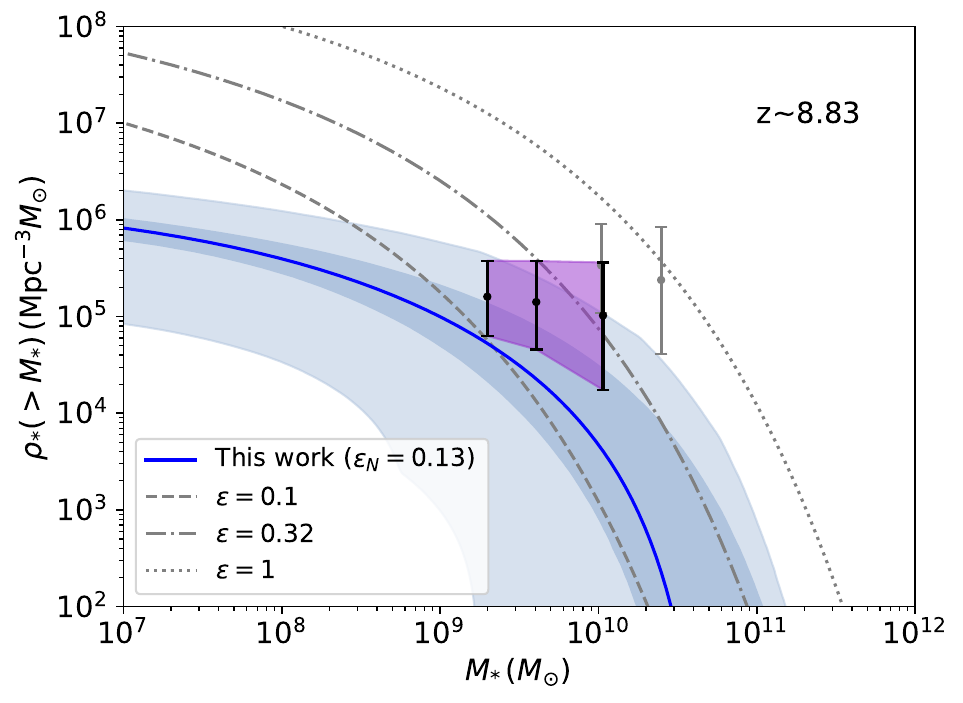}
		\hspace{0mm}}
	
	\caption{\small The comoving stellar mass density contained within galaxies that are more massive than $M_{*}$ at $z\sim7.5$ and $z\sim8.9$. The gray data points and lines are reproduced from Fig.~2 in \citet{2023NatAs.tmp...77B}. The solid blue lines are the best-fit $\rho_{*} (>M_{*})$ considering the best-fit result of SFE at $z=8$ and $z=9$, respectively. The slightly dark blue regions represent the 90\% credible regions of the estimated $\rho_{*}$, while the light blue regions represent the additional contributions of cosmic variance (CV). The purple regions show the updated observations from \citet{2023Natur.616..266L}. Particularly, in the left panel, the orange region represents the cumulative stellar mass density excluding the single/most massive galaxy with $\log (M_{*}/M_{\odot}) \sim 10.89$, suggesting that such a single galaxy dominates and magnifies $\rho_{*} (>M_{*})$ significantly. This most massive galaxy has not been confirmed or excluded by spectroscopic evidence because of no data up to now. }
	\label{fig:2}
\end{figure*}

To examine the reliability of SFE that we inferred, we fit UV LFs with the observations \citep{2013MNRAS.432.2696M, 2015ApJ...810...71F, 2021AJ....162...47B, 2022ApJS..259...20H} at the redshift range of $4\le z \le 7$ at first. The results are shown in \autoref{fig:1}. At the bright end, data points have an obvious excess compared with the fitting UV LFs, which could be attributed to the extra contribution of AGNs or significant uncertainties in the data, resulting in reduced weight for the bright end during the fitting process. As \citet{2022ApJS..259...20H} illustrated in their Fig.~19,  $\epsilon_{\rm N}$ ranges from $\sim 0.06$ to $\sim 0.6$ at $4\le z\le 7$ in different works \citep{2015ApJ...813...21M,2018PASJ...70S..11H,2018MNRAS.477.1822M,2018ApJ...868...92T}, indicating a conflicting evolution trend of SFE with increasing redshift. The question of whether SFE truly increases, decreases, or remains constant at high redshifts remains unclear. For higher redshifts ($8\le z \le 9$), the fitted UV LFs are displayed in the bottom panel of \autoref{fig:1}. The observed data points, which were fitted with \autoref{eq:4}, are derived from the latest JWST sources \citep{2023arXiv230413721A,2023MNRAS.523.1009B,2023MNRAS.518.6011D,2023ApJS..265....5H,2023ApJ...946L..35M,2023arXiv230202429P}. The magenta data points \citep{2022MNRAS.515.5790L,2022ApJ...928...52F} with error bars were not included in the fitting process, as it could be associated with a potential overdensity in the Extended Groth Strip (EGS) field, leading to an excess at the bright end that can be explained by fiducial models without dust attenuation. Our results match these bright sources at $z=9$ well but still exhibit a deviation at $z=8$. Intriguingly, the best-fit values of  $\epsilon_{N}$ demonstrate a rising-falling trend with increasing redshifts and reached a minimum at $z=9$. All of the best fits of SFE fall within a reasonable range and do not indicate exceptionally high values, which are consistent with \citet{2023arXiv230411911P} and \citet{2023arXiv230517959Q}.

Following \autoref{eq:6}, the cumulative stellar mass densities at $7<z<8.5$ and $8.5<z<10$ are shown in \autoref{fig:2}. In contrast to the analysis of \citet{2023Natur.616..266L}, the update observations exclude two galaxies, 13050 and 35300, which are not very-high-redshift candidates any longer. These two galaxies are assigned to separate bins opportunely, resulting in a decrease in $\rho_{*}$. Particularly at $z\sim9$, the exclusion of the most massive galaxy candidate leads to a significant decline in $\rho_{}(>M_{},z=9)$, as it 
determines the maximum stellar mass and dominates the cumulative stellar mass ($>M_{*}$) at lower stellar mass $M_{*}$. Hence, the observations are in agreement with the model well when considering the effect of CV. At $z\sim7.5$, a similar decrease does not occur due to the substantial contribution of the most massive galaxy candidate with $\log M_{*} \sim 10.89 M_{\odot}$, which has not been confirmed or excluded by spectroscopic analysis. Therefore, in the left panel, we show cumulative stellar mass density without the most massive one and find the model can explain the ``observations" well.

\begin{figure*}[ht!]
	\centering
	\subfigure{
		\includegraphics[width=0.45\textwidth]{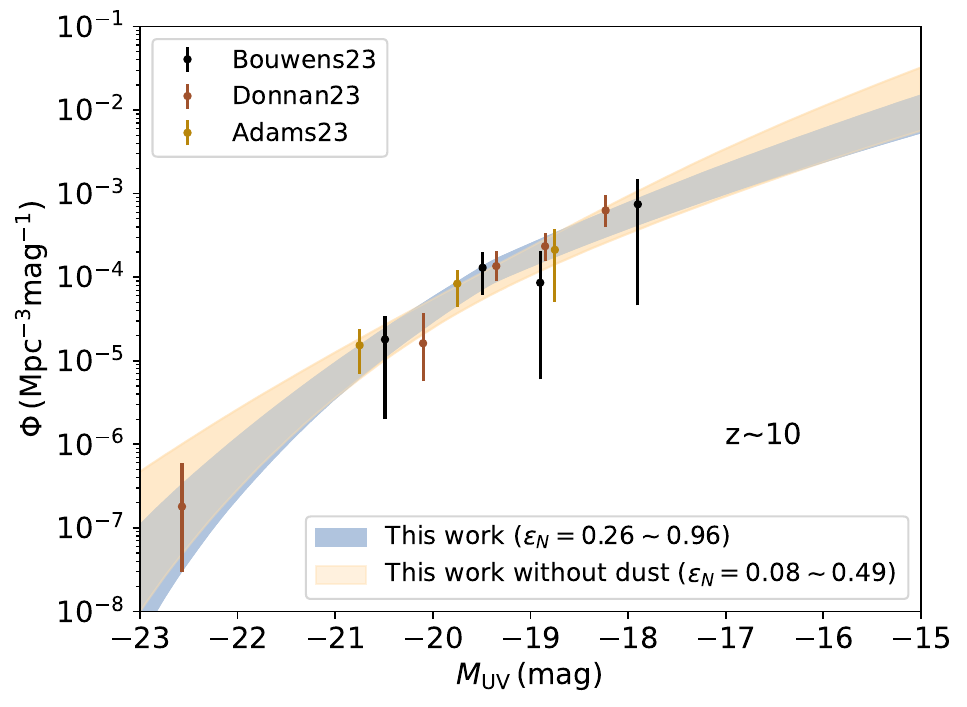}
		\hspace{12mm}}
	\subfigure{
		\includegraphics[width=0.45\textwidth]{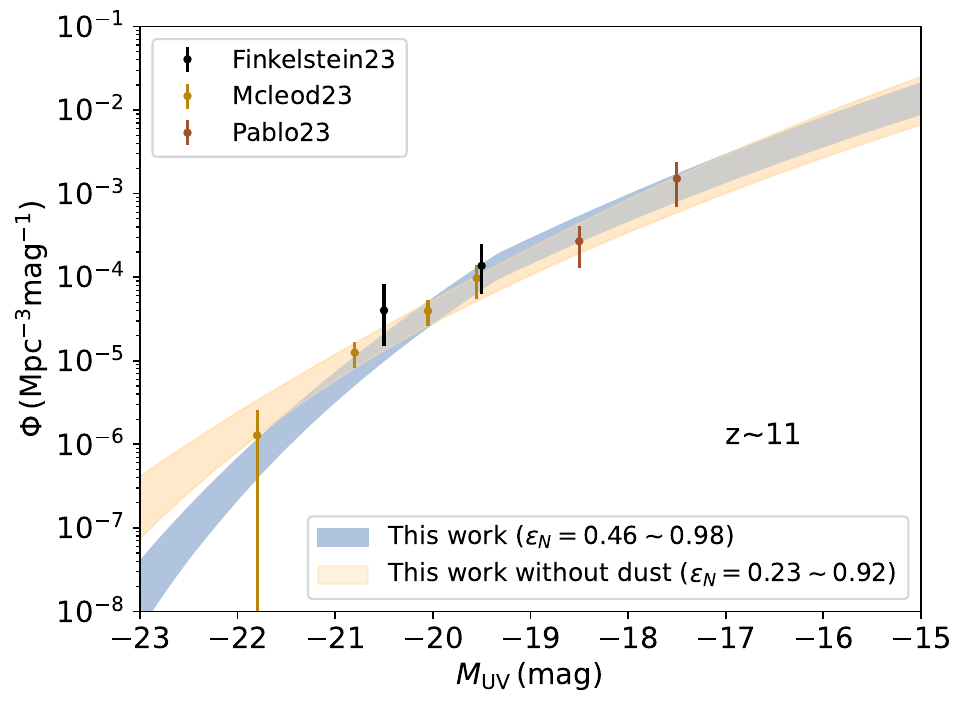}}
	\subfigure{
		\includegraphics[width=0.45\textwidth]{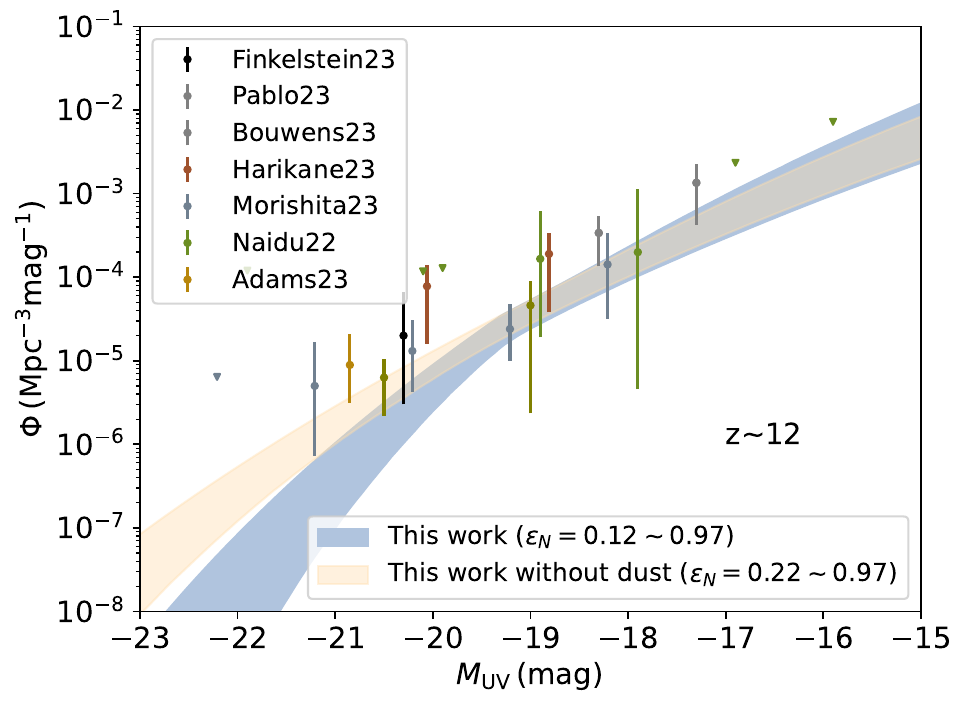}
		\hspace{12mm}}
	\subfigure{
		\includegraphics[width=0.45\textwidth]{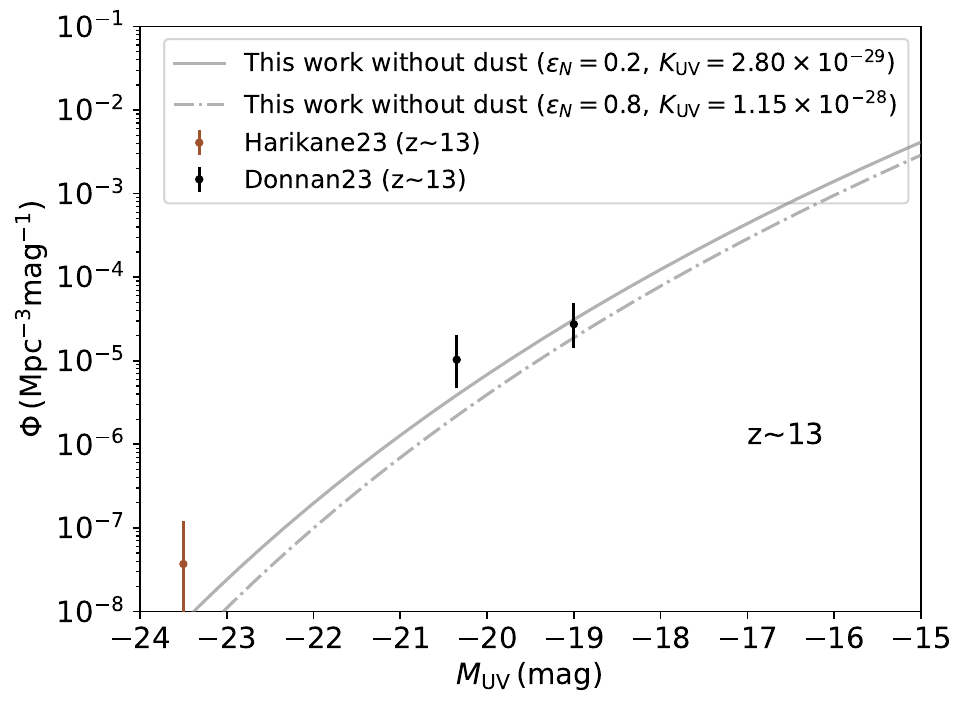}}

	\caption{\small The fit to the observations of UV LFs at the redshift range of $10\le z\le 13$. The selected JWST galaxies are taken from \cite{2022ApJ...940L..55F,2022ApJ...940L..14N, 2023arXiv230413721A, 2023MNRAS.523.1009B, 2023MNRAS.523.1036B, 2023MNRAS.518.6011D, 2023ApJ...946L..13F, 2023ApJS..265....5H, 2023arXiv230414469M, 2023ApJ...946L..35M, 2023arXiv230202429P}. The blue and yellow regions represent the corresponding $90\%$ credible regions of the estimated LFs with and without dust attenuation. The solid, dashed-gray lines represent qualitative cases of ${\mathcal K}_{\rm UV}$ and $\epsilon_{\rm N}$, where the values of $\beta$ and $\gamma$ are fixed at the best fitting values. The observational data points are marked in circles with error bars, and the upper limits are shown in lower triangles.}
	\label{fig:3}
\end{figure*}

\begin{figure}[ht!]
	\centering
	\subfigure{
		\includegraphics[width=0.45\textwidth]{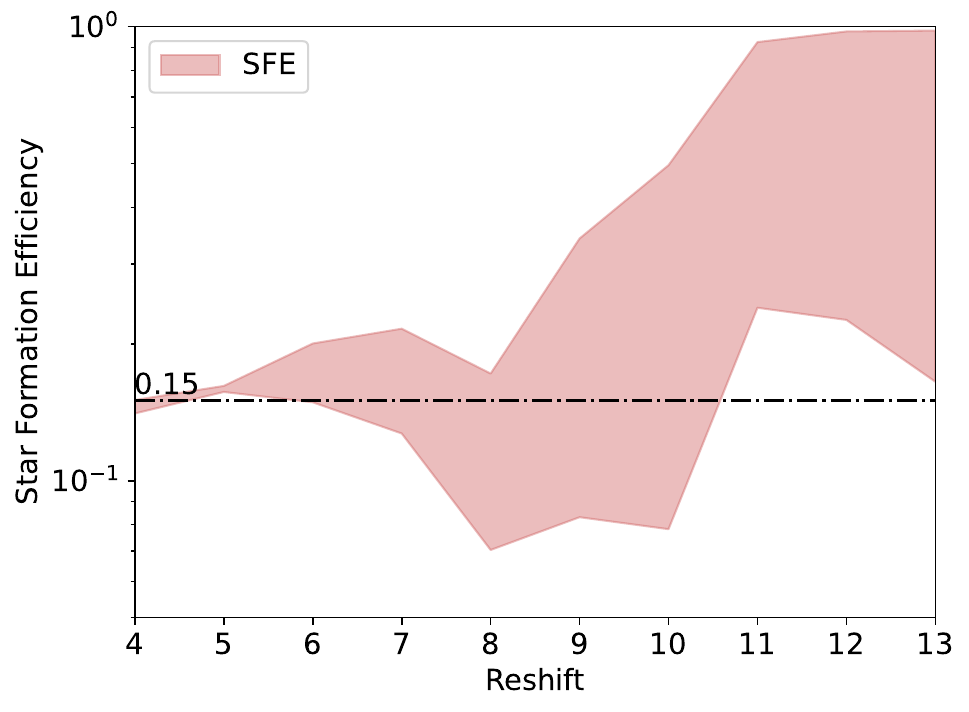}}
	
	\caption{\small The evolution of peak SFE with redshift. The shaded pink region represents the range of maximum SFE values with a $68\%$ credible uncertainty at each redshift. At $z\ge$10, the estimation of SFE is performed without considering the impact of dust attenuation.}
	\label{fig:4}
\end{figure}

At $z\ge10$, whether these bright-UV galaxies can form through the DM-driven channel is still under debate \citep{2022ApJ...939L..31H, 2023arXiv230413755M}. In consequence, we fit UV LFs in two cases, with and without dust attenuation. The observations are derived from JWST sources \citep{2022ApJ...940L..55F, 2022ApJ...940L..14N, 2023arXiv230413721A, 2023MNRAS.523.1009B, 2023MNRAS.523.1036B, 2023MNRAS.518.6011D, 2023ApJ...946L..13F, 2023ApJS..265....5H, 2023arXiv230414469M, 2023ApJ...946L..35M, 2023arXiv230202429P}. Here, our main objective is not to search for the best fitting LF but the estimated range of SEF. Shown in \autoref{fig:3}, the required SFE becomes larger and approaches values that are considered implausible ($\sim 1$) when redshift increases, regardless of whether dust attenuation is taken into account or not. The complete trend of SFE with increasing redshift is illustrated in \autoref{fig:4}. Considering the correlation between SFE and halo mass, we plot the maximum value of SFE at each redshift. In order to provide a conservative estimate, we present the inferred SFE with dust attenuation at $z<10$, and without dust attenuation at $z\ge10$. Our analysis reveals a distinct decline in SFE between two peak epochs, suggesting a potential iterative process in the stellar population. 
Furthermore, all of the posterior distributions of these scenarios (from $z=4$ to $z=12$) are presented in \autoref{sec:7}. The best-fit values of $\epsilon_{\rm N}$, $\beta$, and $\gamma$ are summarized in \autoref{Tab:2}.

At $z\sim13$, we plot UV LFs in some special cases because the data points are limited. Nonetheless,  it is worth noting that the value of ${\mathcal K}_{\rm UV} = 1.15 \times 10^{-28}$ discussed from \citet{2014ARA&A..52..415M} is applicable at $z=8$ and may be lower for higher redshifts since the conversion factor ${\mathcal K}_{\rm UV}$ is sensitive to the metallicity, IMF, and the age of the stellar population. Thus, we calculate LFs using ${\mathcal K}_{\rm UV} = 2.80 \times 10^{-29}$ \citep{2022ApJ...938L..10I}, which is built for metal-free Population III stars with an extremely top-heavy IMF (Salpeter IMF with 50-500 $M_{\odot}$). This low ${\mathcal K}_{\rm UV}$ significantly brightens high-redshift galaxies and leads to distinct enhancements in the UV LFs. We find that the classical ranges of parameters are difficult to explain these bright observations at $z\ge11$, suggesting the possible contribution of Population III stars. Besides, modifying the transfer function in \autoref{eq:1} can enhance the UV LFs within general parameter space \citep[e.g.,][]{2023arXiv230604684P}. At $z\sim 13$, data points can be well fitted with ${\mathcal K}_{\rm UV}=2.80\times 10^{-29}$, providing a plausible explanation for the observed bright UV LFs. Here, we do not discuss the observations at $z\sim16$ \citep{2023MNRAS.523.1009B, 2023ApJS..265....5H}, since the sole high-redshift candidate (CEERS 13256) at $z\sim16$ has been spectroscopically confirmed at $z=4.912\pm 0.001$ \citep{2023arXiv230315431A}.

\section{Conclusions and discussions}\label{sec:4}
Our main purpose is to investigate whether there is a significant discrepancy between the classical galaxy evolution under the $\Lambda$CDM framework and several high-redshift galaxies observed by JWST currently. Particularly, we focus on the cumulative stellar mass density derived from these galaxies, which has been verified has a large deviation (an illogical SFE value) with the base $\Lambda$CDM model. Additionally, the observations of overbright UV luminosity functions (LFs) at redshifts $z\ge10$ have raised considerable debate. By fitting the observations with an analytical UV LFs model, we obtain proper parameter spaces that can explain these anomalies across a wide range of redshifts. 

The analysis of the UV LFs at redshifts $z=7-9$ reveals that a high SFE is not required to explain the observed excess beyond the $\Lambda$CDM model. As \citet{2023MNRAS.tmpL.103P} and \cite{2023arXiv230507049S} pointed out that any modifications to $\Lambda$CDM model that are developed to explain such ultramassive galaxies in the early Universe will be in conflict with HST observations. Similarly, our findings indicate no evidence of modified gravity manifesting in the UV LFs at the range of redshift $z=7-9$. Taking into account the uncertainty from cosmic variance, the exceeding cumulative stellar mass density can be well explained by our inferred SFE at $z\sim9$. Similarly, the observations at $z\sim8$ also align with our expectations when excluding the most massive candidate galaxy. Actually, the sole evidence for departures from the $\Lambda$CDM model is based on the detection of a massive galaxy with a stellar mass of $\log M_{*} = 10.89 M_{\odot}$ at $z=7.48$. Further spectroscopic confirmation is required to accurately determine its stellar mass and redshift. As \citet{2023arXiv230514418B} pointed out,  both a dusty galaxy and an obscured AGN can provide similar spectral energy distribution (SED) fits but exhibit different stellar properties. The former is characterized by a high stellar mass ($\log M_{}/M_{\odot} \sim 10$) and significant dust obscuration ($A_{\rm v} \ge 3$ mag), while the latter exhibits lower stellar mass ($\log M_{}/M_{\odot} \sim 7.5$) and minimal obscuration ($A_{\rm v} \sim 0$ mag). This perspective has been supported by the confusion between a dusty AGN ($z\sim5.62$) and high-redshift galaxy CEERS 3210 ($z\sim8.62$) \citep{2023arXiv230200012K}. To confirm such overdensity, future observations should expand the search field to reduce the effects of CV. We conclude that it is premature to consider the single massive galaxy detected by JWST as conclusive evidence beyond the $\Lambda$CDM model.

Additionally, our analysis suggests a slight declining-increasing tendency for the SFE with increasing redshift, although the detected data points have large uncertainties at higher redshift $z\ge10$ and the accuracy of the estimated SFE is not as reliable as at lower redshift. Even when considering the degeneracy between dust attenuation and SFE, where high attenuation would decrease the observed UV LFs while high SFE would increase it, we still observe an increasing trend in SFE at $z\ge10$. The decreasing trend of SFE from $z\sim 7$ to $z\sim 4$ in \autoref{fig:4} is similar to the redshift dependence of SFE used in cosmological simulations for star formation, which follows the equation $\epsilon(z)=\epsilon _0+\epsilon _z \times(\frac{z}{z+1})$, where $\epsilon_0$ and $\epsilon_z$ are constants and z is the redshift \citep{2018MNRAS.477.1822M}. 
The weak dip in SFE at redshift $z\sim8-9$ in \autoref{fig:4}, as well as the decreasing trend from $z\sim11$ to $z\sim8$, may be attributed to feedback from early star formation activities of Pop III stars \citep{2009Natur.459...49B,2021MNRAS.506.5247L,2022arXiv221204476W,2023arXiv230600953M}. The rapid supernovae explosions driven by massive Population III stars in the early universe can significantly impact their surrounding interstellar medium \citep{2008IAUS..255...24T,Ma:2023oko}, leading to reheating that prevent subsequent star formation at high redshift around $z\sim10$. The SFE may increase after the completion of Population III activities at lower redshifts around $z\sim7$. As this possible feedback effect extends over larger volumes during cosmic evolution, it causes a decrease in SFE from high redshifts to low redshifts.
The observed dip in SFE at around $z\sim8$ will be crucially constrained or confirmed by future observations with JWST. At $z\sim13$, high SFE and absence of dust attenuation are difficult to explain these overabundance of bright sources. Therefore, Population III star might show nonnegligible contribution to their brightness, as \citet{2023arXiv230504944T}, \citet{2023MNRAS.522.3809V} and \citet{2023arXiv230404348Y} suggested. 

In conclusion, it is premature to claim that JWST has detected something beyond $\Lambda$CDM model. The derived stellar masses from different SED fitting algorithms can vary by almost 1 order of magnitude, as highlighted by \citet{2023ApJ...945L..21V}. Specifically, previous controversies regarding the confusion between Lyman break and Balmer break galaxies at $z\sim5$ could lead to erroneous estimations, resulting in dusty/quenched galaxies appearing strikingly luminous at $z\sim17$ \citep{2022arXiv220802794N}, which has been confirmed in \citet{2023arXiv230315431A} with strong emission lines. Therefore, it is crucial to exercise caution when discussing whether the $\Lambda$CDM model is inadequate for understanding the early universe. Further investigations and the accumulation of more robust data are necessary to validate and refine these initial findings.

\begin{acknowledgements}
\section{Acknowledgements}
We thank the anonymous referee for helpful comments and suggestions. We thank Shao-Peng Tang for his help in developing efficient codes, and Shi-Jie Gao for the helpful discussions. This work is supported in part by NSFC under grants of No. 11921003 and No. 12233011; G.W Yuan is supported by the China Postdoctoral Science Foundation under grant No. 2023TQ0355.

Software: {\tt Pymultinest} (\citet{2016ascl.soft06005B}, version 2.11, \url{https://pypi.org/project/pymultinest/}), {\tt HMFcalc} (\citet{2014ascl.soft12006M}), \url{https://github.com/halomod/hmf/}
\end{acknowledgements}

\appendix

\section{Comparisons with different prior assumptions}\label{sec:6}
\begin{figure*}[ht!]
	\centering
	\subfigure{
		\includegraphics[width=0.45\textwidth]{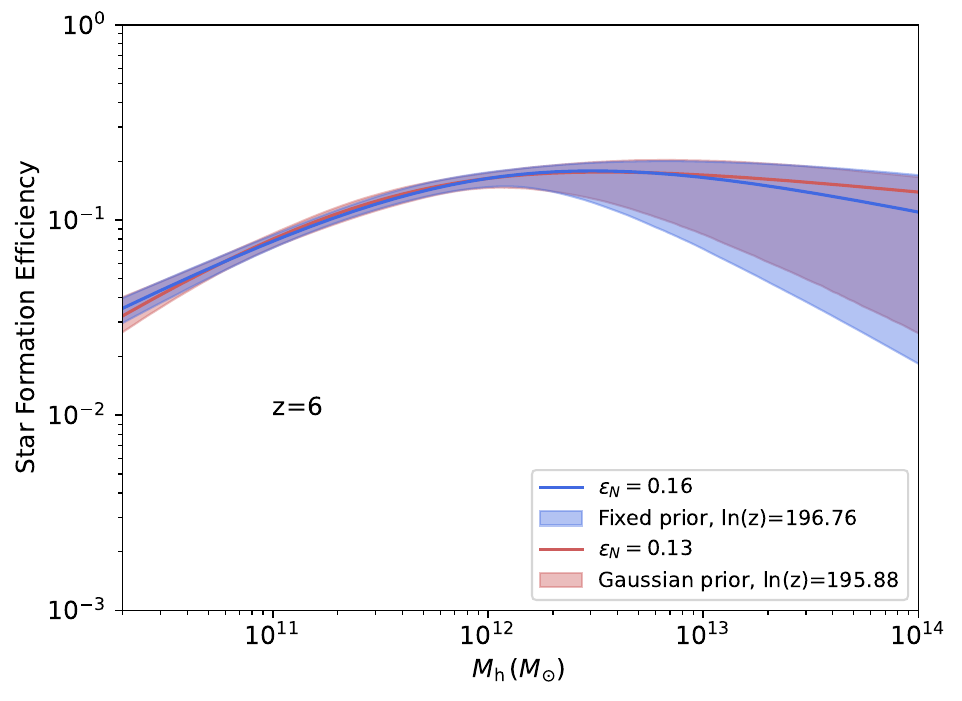}
		\hspace{12mm}}
	\subfigure{
		\includegraphics[width=0.45\textwidth]{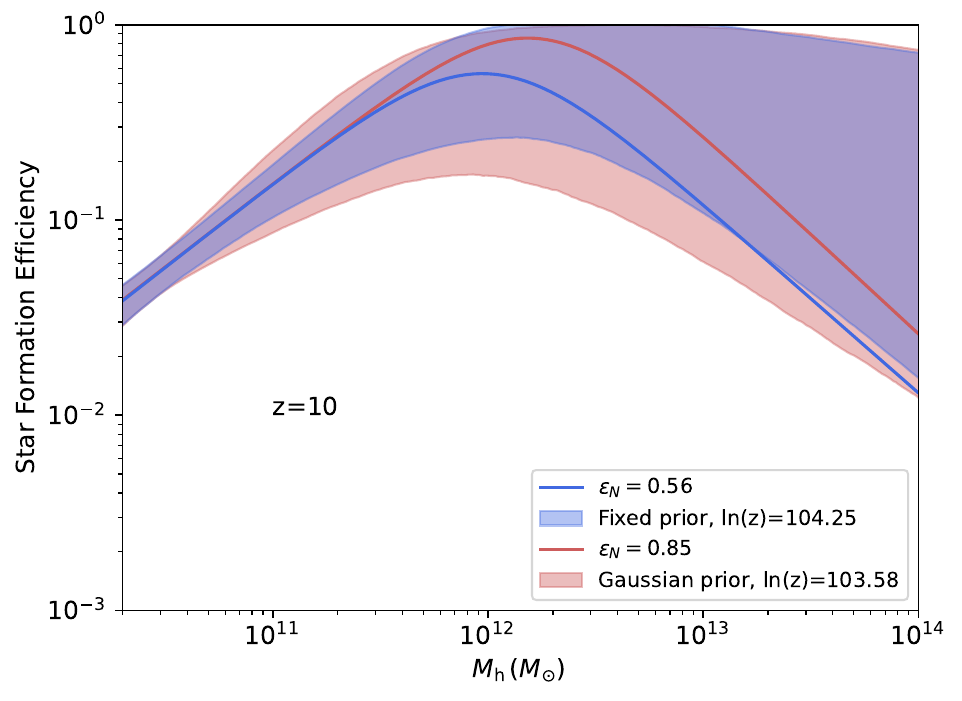}}
	\caption{\small The estimated SFEs with a fixed prior and a Gaussian prior of $M_1$ are shown in red and blue, respectively. The solid lines and colored regions represent the best-fit value and 90\% credible level of SFE. The similar logarithms of Bayes evidence $\ln(Z)$ suggest that it is inconclusive to distinguish the better fitting model.}
	\label{fig:7}
\end{figure*}

The parameter $M_1$ of SFE is fixed as a constant in our previous discussion. In \autoref{fig:7} we show that the resulting profile of $f_* - M_{\rm h}$ is similar if the prior of $\log(M_1)$ follows the Gaussian distribution with the parameters of $\mu=12 M_{\odot}$ and $\sigma=1M_{\odot}$. The redshifts of $z=6$ and $z=10$ have been taken for illustration.

\section{The posterior distributions of the parameters of SFE model}\label{sec:7}

\begin{figure*}[ht!]
	\centering
	\subfigure{
		\includegraphics[width=0.45\textwidth]{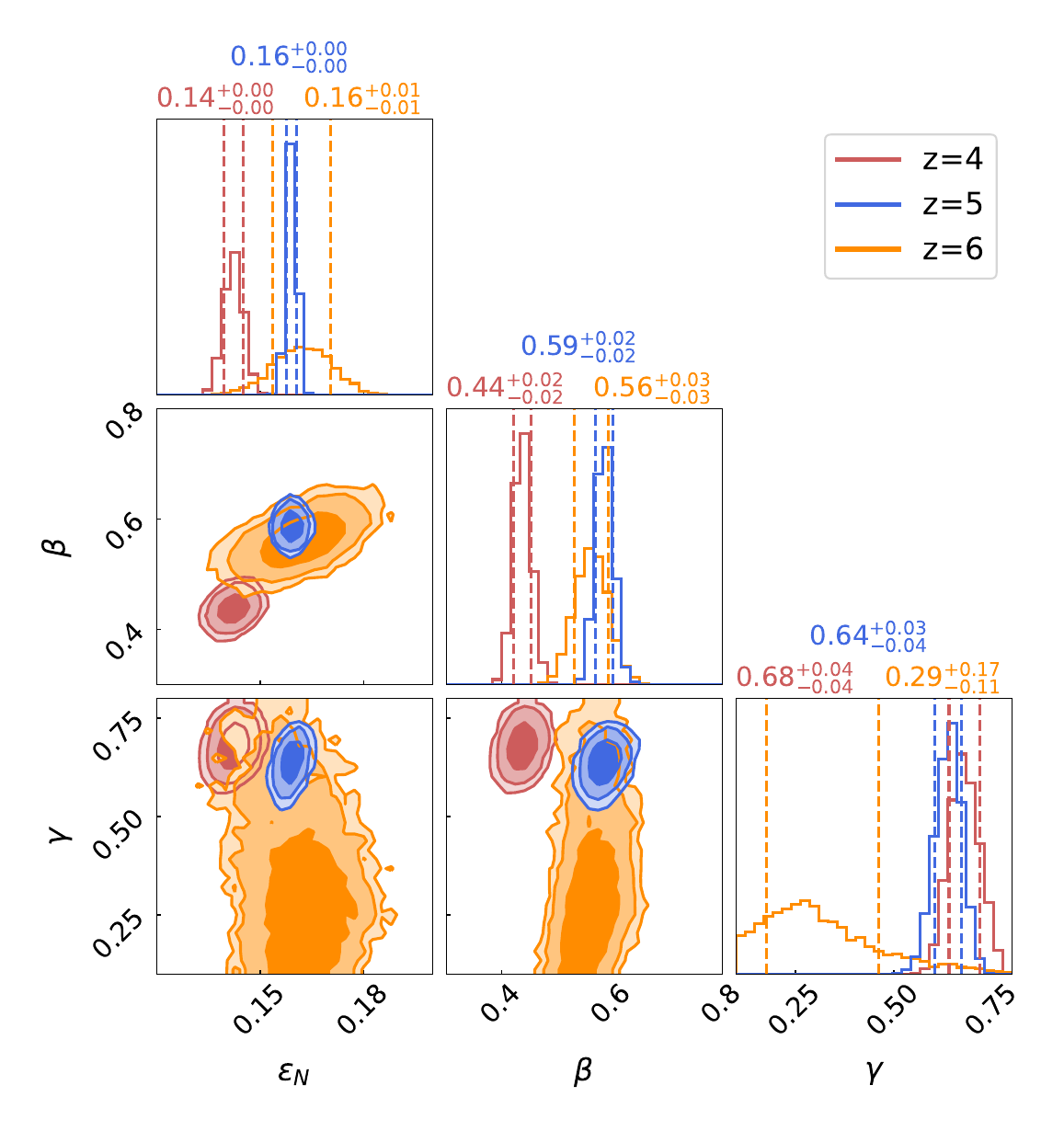}
		\hspace{12mm}}
	\subfigure{
		\includegraphics[width=0.45\textwidth]{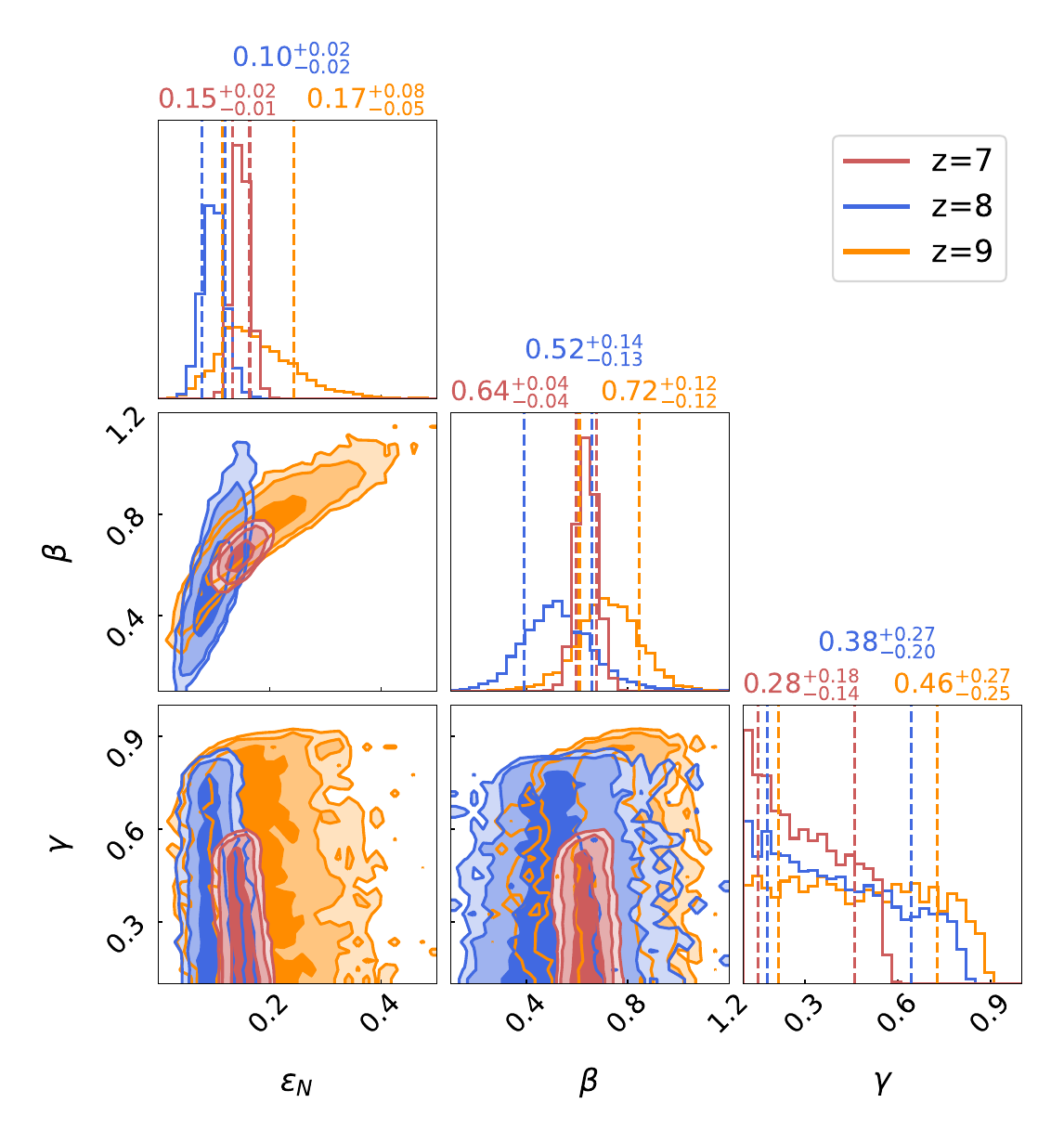}}
	\caption{\small Posterior distributions of three SFE parameters for the best-fit UV LFs presented in \autoref{fig:1}. The results for different redshifts are shown in red, blue, and orange, respectively. The contours are at the 68\%, 95\%, and 99\% credible levels. The values correspond to the 68\% credible level. {It should be noticed that the parameter ranges are employed distinct for better visual effect.}}
	\label{fig:5}
\end{figure*}

\begin{figure*}[ht!]
	\centering
	\subfigure{
		\includegraphics[width=0.45\textwidth]{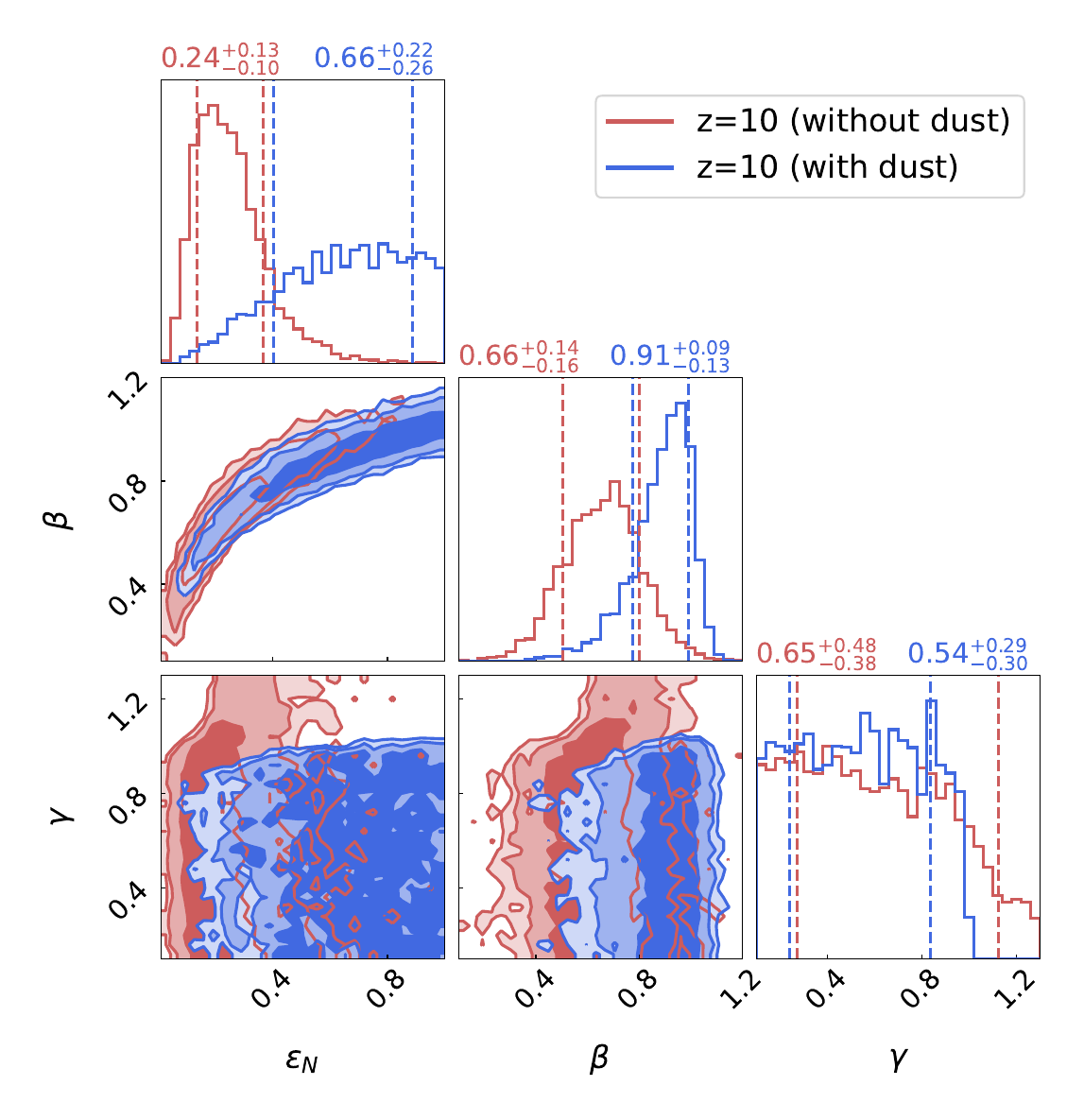}
		\hspace{12mm}}
	\subfigure{
		\includegraphics[width=0.45\textwidth]{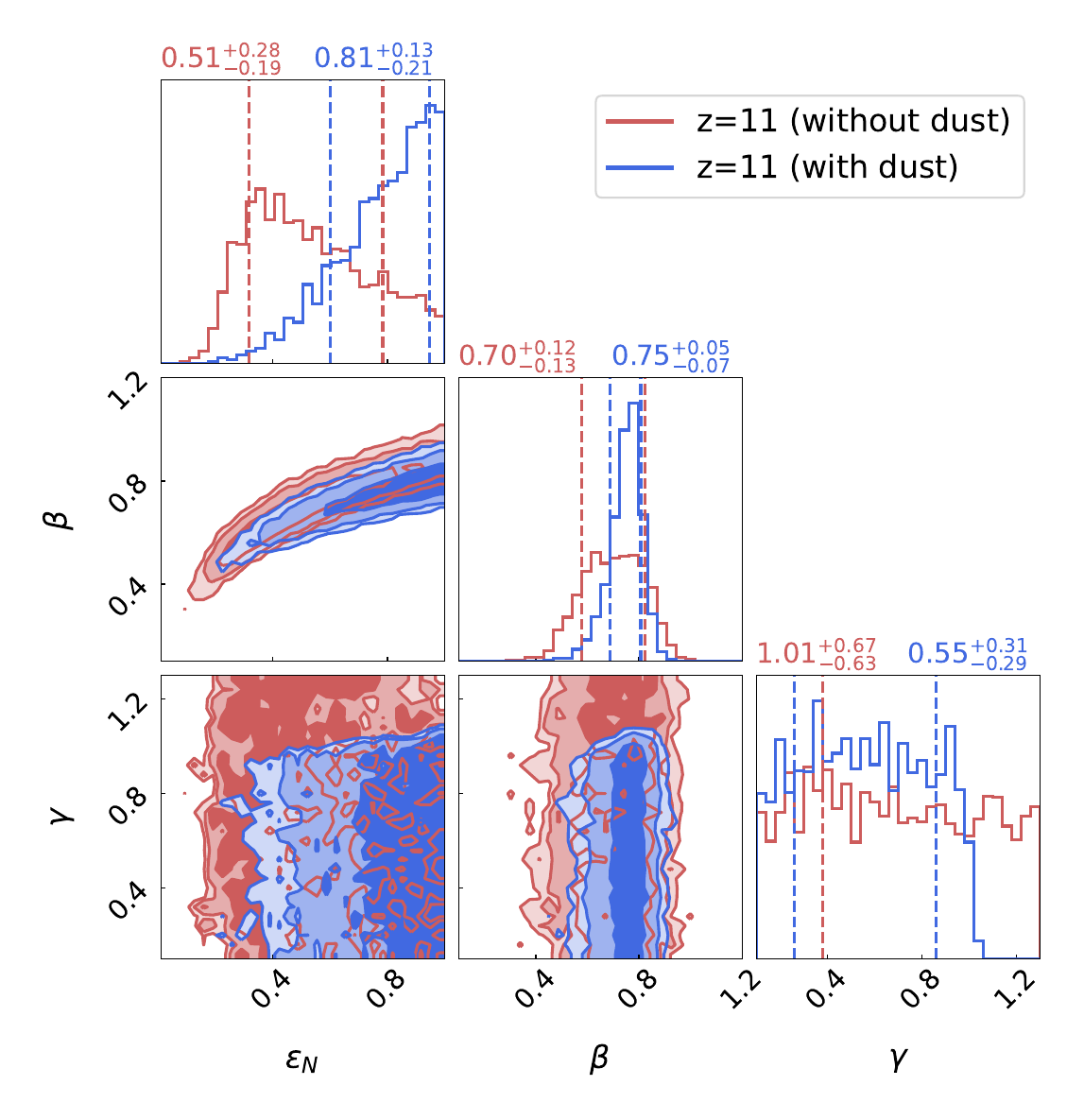}}
	\subfigure{
		\includegraphics[width=0.45\textwidth]{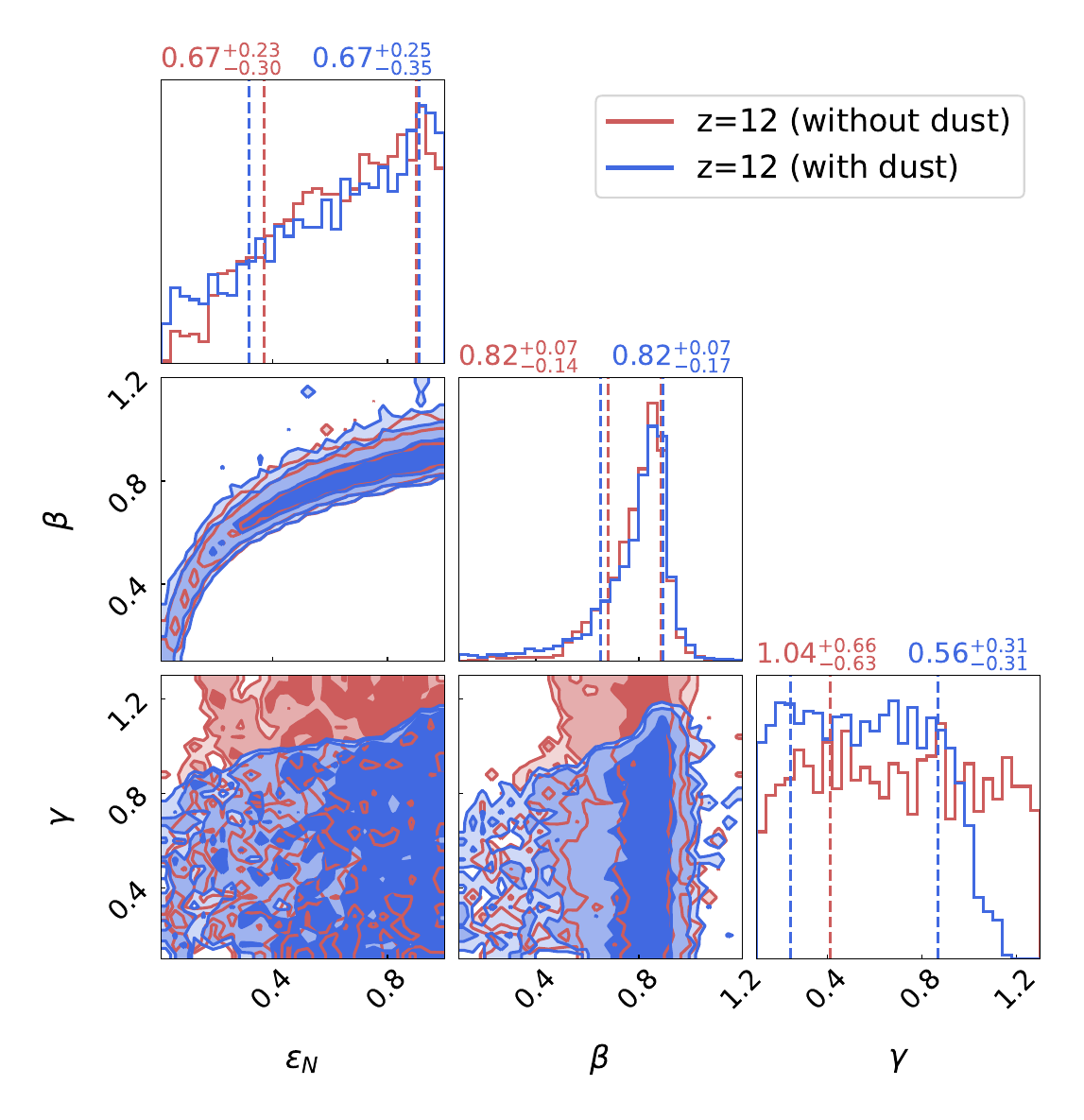}
		\hspace{12mm}}
	
	\caption{\small Posterior distributions of three SFE parameters for the best-fit UV LFs presented in \autoref{fig:3}. The results are displayed in blue and red, differentiating between cases with and without consideration of dust attenuation. The contours are at the 68\%, 95\%, and 99\% credible level. The values correspond to the 68\% credible level.}
	\label{fig:6}
\end{figure*}

Here, we present the posterior distributions of the SFE parameters mentioned in \autoref{sec:3}. As an extra supplement for UV LFs fitting, \autoref{fig:5} and \autoref{fig:6} show the scenarios in the redshift range of $4\le z \le 12$ with dust attenuation, including the extra comparisons without dust attenuation at high redshift ($z\ge10$). Furthermore, \autoref{Tab:2} shows the best-fit values of the SFE parameters. Because of the quality and quantity of the observed data points, the estimated uncertainties of SFE parameters become larger at higher redshifts. In addition, $\gamma$ governs the variation of SFE at high mass. At high redshift, the density of high-mass DM halo is very low, hence the change of $\gamma$ cannot effectively modify the bright end of the UV LFs. That is why the $\gamma$ can not be well constrained at high redshift.

\begin{table*}[ht!]
\begin{ruledtabular}
\centering
\caption{The best-fit Values and Posterior Results of the SFE parameters}
\label{Tab:2}
\begin{tabular}{c|ccc|ccc|c}
\multirow{2}{*}{Redshift} & &Best-fit Values & &\multicolumn{3}{c|}{Posterior Results at 68\% Credible Level} &\multirow{2}{*}{$\ln(Z)$} \\
&$\epsilon_{\rm N}$ &$\beta$ &$\gamma$ &$\epsilon_{\rm N}$ &$\beta$ &$\gamma$   \\ \hline     
4 &0.14 &0.44 &0.67 &$0.14^{+0.003}_{-0.002}$ &$0.44^{+0.02}_{-0.02}$ &$0.68^{+0.04}_{-0.04}$  &376.69\\
5 &0.16 &0.59 &0.64 &$0.16^{+0.001}_{-0.001}$ &$0.59^{+0.02}_{-0.02}$ &$0.64^{+0.03}_{-0.04}$ &253.30 \\
6 &0.16 &0.56 &0.23 &$0.16^{+0.01}_{-0.01}$ &$0.56^{+0.03}_{-0.03}$ &$0.29^{+0.17}_{-0.11}$ &196.76\\
7 &0.16 &0.64 &0.11 &$0.15^{+0.02}_{-0.01}$ &$0.64^{+0.04}_{-0.04}$ &$0.28^{+0.18}_{-0.14}$ &279.49\\
8 &0.11 &0.46 &0.10 &$0.10^{+0.02}_{-0.02}$ &$0.52^{+0.14}_{-0.13}$ &$0.38^{+0.27}_{-0.20}$ &85.44\\
9 &0.13 &0.65 &0.80 &$0.17^{+0.08}_{-0.05}$ &$0.72^{+0.12}_{-0.12}$ &$0.46^{+0.27}_{-0.25}$ &143.52\\
10 &0.56 &0.86 &0.97 &$0.66^{+0.22}_{-0.26}$ &$0.91^{+0.09}_{-0.13}$ &$0.54^{+0.29}_{-0.30}$ &104.25\\
10\textsuperscript{a} &0.15 &0.53 &0.39 &$0.24^{+0.13}_{-0.10}$ &$0.66^{+0.14}_{-0.16}$ &$0.65^{+0.48}_{-0.38}$  &103.12\\
11 &0.98 &0.80 &0.96 &$0.81^{+0.13}_{-0.21}$ &$0.75^{+0.05}_{-0.07}$ &$0.55^{+0.31}_{-0.29}$ &69.49\\
11\textsuperscript{a}&0.40 &0.61 &0.14 &$0.51^{+0.28}_{-0.19}$ &$0.70^{+0.12}_{-0.13}$ &$1.01^{+0.67}_{-0.63}$ &70.56\\
12 &0.99 &0.89 &1.13 &$0.67^{+0.25}_{-0.35}$ &$0.82^{+0.07}_{-0.17}$ &$0.56^{+0.31}_{-0.31}$ &127.37\\
12\textsuperscript{a}&0.99 &0.89 &1.90 &$0.67^{+0.23}_{-0.30}$ &$0.82^{+0.07}_{-0.14}$ &$1.04^{+0.66}_{-0.63}$ &129.27\\
\end{tabular}
\end{ruledtabular}
\begin{tablenotes}
 \item[a] \textsuperscript{a} The values in these rows represent the fitting results without considering the attenuation effect.
\end{tablenotes}
\end{table*}

\clearpage
\bibliography{ref}
\bibliographystyle{aasjournal}
\end{document}